\documentclass[apjl]{emulateapj}

\def\mt{ }
\def\vv{ }

\def\be{\begin{equation}}
\def\en{\end{equation}}

 \slugcomment{Submitted to
 \textit{ApJ Letters}, 15 Mar., 2011}

\begin{document}

\title{Spectral Evolution of the September 2010 gamma-ray flare from the Crab Nebula}




\author{V.~Vittorini\altaffilmark{1}, M.~Tavani\altaffilmark{1,2},
G.~Pucella\altaffilmark{3},
E. Striani\altaffilmark{2}, I. Donnarumma\altaffilmark{1},
P.~Caraveo\altaffilmark{4}, A. Giuliani\altaffilmark{4},
S. Mereghetti\altaffilmark{4}, A. Pellizzoni\altaffilmark{5},
A. Trois\altaffilmark{1,5 }, A. Ferrari\altaffilmark{6,7}
G.~Barbiellini\altaffilmark{8}, A.~Bulgarelli\altaffilmark{9},
P.~W.~Cattaneo\altaffilmark{10}, S.~Colafrancesco\altaffilmark{11},
E.~Del Monte\altaffilmark{1}, Y.~Evangelista\altaffilmark{1},
F.~Lazzarotto\altaffilmark{1}, L.~Pacciani\altaffilmark{1},
M.~Pilia\altaffilmark{12},  C.~Pittori\altaffilmark{11}}

\altaffiltext{1} {INAF/IASF-Roma, I-00133 Roma, Italy}
\altaffiltext{2} {Dip. di Fisica, Univ. Tor Vergata, I-00133 Roma,
Italy} \altaffiltext{3}{ENEA Frascati, via Enrico Fermi 45,
I-00044 Frascati(Roma), Italy} \altaffiltext{4} {INAF/IASF-Milano,
I-20133 Milano, Italy} \altaffiltext{5} {INAF-Osservatorio
Astronomico di Cagliari, localita' Poggio dei Pini, strada 54,
I-09012 Capoterra, Italy} \altaffiltext{6} {Dip. Fisica,
Universit\'a di Torino, Turin, Italy} \altaffiltext{7} {Consorzio
Interuniversitario Fisica Spaziale, Villa Gualino, I-10133 Torino,
Italy} \altaffiltext{8} {Dip. Fisica and INFN Trieste, I-34127
Trieste} \altaffiltext{9} {INAF/IASF-Bologna, I-40129 Bologna,
Italy} \altaffiltext{10} {INFN-Pavia, I-27100 Pavia, Italy}
\altaffiltext{11} {ASI Science Data Center, I-00044}
\altaffiltext{12} {Dipartimento di Fisica, Universit\'a
dell'Insubria, Via Valleggio 11, I-22100 Como, Italy}

\begin{abstract}

Strong gamma-ray flares from the Crab Nebula have been recently
discovered by AGILE and confirmed by Fermi-LAT.
We study here the spectral evolution in the gamma-ray energy range
above 50 MeV of the September 2010 flare that was simultaneously
detected by AGILE and Fermi-LAT. We revisit the AGILE spectral
data, and present an emission model based on rapid (within 1 day)
acceleration followed by synchrotron cooling. We show that this
model successfully explains both the published AGILE and Fermi-LAT
spectral data showing a rapid rise and a decay within 2-3 days.
Our analysis constrains the acceleration timescale and mechanism,
the local magnetic field, and the particle distribution function. 
The combination of very rapid acceleration,
emission well above 100 MeV, and the spectral evolution consistent
with synchrotron cooling contradicts the idealized scenario
predicting an exponential cutoff at photon energies above 100 MeV.
We also consider a variation of our model based on even shorter
acceleration and decay timescales which can be consistent with the
published averaged properties.

\end{abstract}



\maketitle


\section{Introduction}

The Crab Nebula  is at the center of the SN1054 supernova remnant
and consists of a rotationally-powered pulsar interacting with a
surrounding nebula through a relativistic particle wind (e.g.,
Hester 2008). The Crab pulsar is quite powerful (of spindown
luminosity $\textit{L$_{PSR}$} $= 5$\cdot$10$^{38}$ erg s$^{-1}$,
and spin period $\textit{P}$ = 33 ms), and is energizing the whole
nebula with its wave/particle output.  The inner nebula shows
distinctive optical and X-ray brightness enhancements (``wisps'',
``knots'', and the ``anvil'' aligned with the pulsar ``jet'')
(Scargle 1969; Hester 1995, 2002, 2008; Weisskopf 2000).
These local
variations have been attributed to enhancements of the synchrotron
emission produced by instabilities and/or shocks in the pulsar
wind outflow. However, when averaged over the whole inner region
(several arcminute across) the Crab Nebula has been considered
essentially stable, and used as a "standard candle" in high-energy
astrophysics.
The Crab Nebula X-ray
continuum and gamma-rays up to $\sim 100 \, $MeV energies are
modelled by synchrotron radiation of accelerated particles in an
average nebular magnetic field $\bar{B} = 200 \, \mu$G (Hester
2002, deJager  et al. 1996, Atoyan \& Aharonian 1996, Meyer et al.
2010).
Emission from GeV to TeV energies is interpreted as inverse
Compton radiation by electrons/positrons scattering CMB and
nebular soft photons (deJager \& Harding 1992, deJager etal. 1996,
Atoyan \& Aharonian 1996, Meyer et al. 2010).

Decades of theoretical modelling of this system (e.g., (Rees \&
Gunn 1974, Kennel \& Coroniti 1984, deJager \& Harding 1992,
deJager etal. 1996, Atoyan \& Aharonian 1996, Arons 2008, Meyer et
al. 2010)
offer the picture of a remarkable nebular system energized by a
MHD pulsar wind interacting with the environment through a
sequence of "shocks" or dissipation features localized at
distances larger than a few times $10^{17}$cm. Efficient particle
acceleration at the pulsar wind termination shock regions is
believed to be occurring either through diffusive processes, e.g.,
\citep{blandford,dejager1,dejager2,atoyan}, shock-drift
acceleration (e.g., Begelman \& Kirk 1990) or ion-mediated
acceleration (e.g., \cite{arons,spitkovsky}). Several diffusive
acceleration models imply acceleration rates of order of the
relativistic electron cyclotron frequency (e.g.,
\cite{dejager1,dejager2,atoyan}). Assuming  equality between the
accelerating electric field and the magnetic field at the
acceleration site and synchrotron cooling in the co-spatial
magnetic field leads to a most cited constraint for the maximum
radiated photon energy (e.g., Aharonian 2004)
\be E_{\gamma, max} \simeq \frac{9}{4}\alpha^{-1} \, m_e \, c^2 \simeq 150 \,
\rm MeV \label{eq-1} \en
with $\alpha = e^2/\hbar \, c$ the fine structure constant, $c$
the speed of light, and $m_e$ the electron's mass. Eq. \ref{eq-1}
applies in a natural way to diffusively accelerated particles, and
$E_{\gamma, max}$ turns out to be independent of the local
magnetic field. According to the assumptions underlying this
formula, emission above 100 MeV would be difficult to sustain in
the Crab Nebula environment. Indeed, the exponential cutoff shown
by the average gamma-ray spectrum in the 10 MeV - 10 GeV range
supports this idealization \cite{dejager1,dejager2}.

However, the recent discovery by the AGILE satellite of a strong
gamma-ray flare above 100 MeV from the Crab Nebula in September
2010 \cite{tavani1,tavani3} and the confirmation by the Fermi-LAT
\cite{buehler,abdo2} substantially change this picture. Three
substantial gamma-ray flaring episodes from the Crab Nebula have
been announced so far (Tavani et al. 2011, hereafter T11; Abdo et
al. 2011, hereafter A11).  The flaring activity was detected
only in the gamma-ray energy range 100 MeV - a few GeV, and it is
attributed to transient nebular unpulsed emission. 
No significant variations were detected in other bands, except for 
the local enhancement in the "anvil" region revealed by high spatial 
resolution observations of HST and \textit{chandra}.

Three features of the September 2010 event are relevant: (1) the
event develops within 3-4 days (whereas the others last about 2
weeks); (2) the gamma-ray risetime appears to be remarkably short,
$\tau \leq 1 \, $day (T11); (3) the flaring gamma-ray spectrum
extends well above the limit of Eq. 1 (T11,A11). A flare
production site in the inner nebula of size \textit{L} $\leq$
10$^{16}$ cm is favored by both the peak isotropic gamma-ray
luminosity \textit{L$_{p}$} $\approx$ 5$\cdot$10$^{35}$ erg
s$^{-1}$ (which implies for a (3-5) $\%$ radiation efficiency that
about (2-3)\% of the total spindown pulsar luminosity is
dissipated at the flaring site) and by the flare risetime of
$\sim$1 day. We noticed that the ``anvil region'' ("knot-2" and
possibly "knot-1") in the Crab Nebula \cite{scargle,hester1} is an
excellent flare site candidate also because of its alignment with
the relativistic pulsar jet (T11).

A number of important theoretical questions are raised by these
detections. However, the published  spectra of the September 2010
event are not homogeneous because of different integration times:
a 2-day timescale for the AGILE data (T11), and a 4-day timescale
for the Fermi-LAT data. The spectral shapes also appear different.
The AGILE data are characterized by a hard curved spectral shape
with peak photon energy $E_p$ of the differential power spectrum
$E_p \simeq 300 \, $MeV (T11, see Fig. 1). On the contrary, the
Fermi-LAT spectrum shows a quasi power-law shape extending up to a
few GeV (A11). Without additional analysis, it is not clear
whether the two datasets are consistent with each other. In any
case, the hardness of the gamma-ray emission and the rapid
spectral evolution challenge the idealized scenario underlying Eq.
\ref{eq-1} \cite{dejager1,dejager2,atoyan}.

The goal of our paper is twofold: (1) investigate the consistency
of the published AGILE and Fermi-LAT spectral data of the Sept.
2010 event by integrating the AGILE data over a 4-day timescale;
(2) study the spectral evolution of a class of synchrotron
emission models based on freshly accelerated particles in the
inner Nebula, and check its validity for both the AGILE and
Fermi-LAT data.

\section{Spectral data analysis}

In order to test whether the AGILE and Fermi-LAT spectra of the
Sept. 2010 event are consistent with each other it is necessary to
consider data with the same integration timescales. In the absence
 Fermi-LAT spectral data on a 2-day timescale, we revisited our
AGILE data and carried out a 4-day integration which overlaps with
the Fermi-LAT interval.

The AGILE 4-day Crab spectrum in the energy range 50 MeV-3 GeV was
obtained by integrating between MJD 55457.38-55461.55. {\vv We
obtained the nebular contribution by subtracting from the total
emission the pulsar contribution corresponding to a flux F($E>
100$ MeV)= $210\pm 30\,$ph$\,$cm$^{-2}$s$^{-1}$ characterized by 
a power-law spectrum with photon index $\Gamma= 2.0\pm 0.1$ in 
the energy range 50 MeV - 3 GeV.}

Fig. \ref{fig-2} shows the result of our additional analysis of
the AGILE data together with the published Fermi-LAT data. The two
data sets are now temporally homogeneous {\mt and appear to be in
agreement} within  the errors. The apparent power-law behavior of
the 4-day spectrum in the energy range 50 MeV - 2 GeV can be
explained by a fast-rise-synchrotron-cooling model (see below).
The solid curve of Fig. \ref{fig-2} shows the result of our
modelling for a 4-day average of the rapidly varying spectrum.

\section{A fast-rise synchrotron cooling  model}

We assume that a fresh population of impulsively accelerated
electrons/positrons is produced in the inner nebula within a
timescale short compared with all other relevant cooling
timescales. The model presented here has general validity and does
not depend on a specific site in the Nebula as long as the general
characteristics of the emission fit our assumptions. We assume an
efficient particle acceleration mechanism\footnote{We leave for
other investigations the crucial issue of explaining the type of
plasma wave turbulence leading to the short acceleration timescale
(1 day or shorter).} that applies simultaneously in one or more
contiguous nebular sites that are subject to plasma instabilities
and/or substantial pulsar wind particle density enhancements. A
fraction of the total volume of the inner nebula is affected by
the flaring instability. Consequently, only a fraction of the
total number of radiating nebular particles contributes to the
flare. This fact follows from the observed short flaring
rise-time. For simplicity, we assume a Doppler factor ${\cal D} =
(1 - \beta \cos \theta)^{-1} \sim 1$; a larger Doppler factor
would imply a smaller particle number $N_e\propto {\cal D}^{-3}$,
a larger emitting region $L\propto {\cal D}$, and smaller
rest-frame particle energies $\gamma\propto {\cal D}^{-1/2}$.

In our analysis we considered different values of the local
magnetic field $B_{loc}$.
 The rapid observed cooling (2-3
 days) for reasonable values of particle energies imply
 that the local magnetic field is substantially enhanced compared
 with $\bar{B}$ (T11, A11). Reconciling the synchrotron cooling timescale $\tau_s
 \sim (8 \cdot 10^{8} \,{\rm sec}) \,  B_{loc}^{-2} \,  \gamma^{-1}$ (where the local
magnetic field $B_{loc}$ is in Gauss, and $\gamma$ is a typical
particle Lorentz factor) with the Sept. 2010 observations implies,
for $\gamma \approx 5\times 10^{9}$ of electrons irradiating in
the GeV range, a local magnetic field  $B_{loc} \simeq 10^{-3}\,
$G that is $\sim$ 5 larger than the nebular average.

Our best modelling assumes an emitting region of size $L = 7
\times 10^{15}\, $cm, and an enhanced local magnetic field
$B_{loc} = 10^{-3} \,$G that we keep constant in our
calculations.
 The acceleration process produces, within a timescale shorter
than any other relevant timescale,
 a particle energy distribution {\vv that we model as a double power-law
 distribution (T11)}
\begin{equation}
\frac{dn}{d\gamma}=\frac{K\gamma_b^{-1}}{(\gamma/\gamma_b)^{p_1}
+(\gamma/\gamma_b)^{p_2} \label{eq-2}}
\end{equation}
\noindent where $n$ is the particle number density.  
The assumption ${\cal D} \sim 1$ together with the 
constraint on $B_{loc}$ implies a break energy $\gamma_b$ around 
$2 \times 10^9$, and a normalization factor $K$ around 
$5\times 10^{-10}$cm$^{-3}$. 
If the gamma-ray flare is related with 
the persistent local enhancement detected in the anvil region 
by HST and \textit{chandra} (T11), we can constrain $p_1=2.1$ and $p_2=2.7$, 
with the particle Lorentz factor 
$\gamma$ ranging from $\gamma_{min}=10^6$ to $\gamma_{max}=7\times 10^{9}$.
The double power-law distribution of  Eq. \ref{eq-2} implies maximal
synchrotron emission between $\gamma_b$
and $\gamma_{max}$ and the total particle number required to explain 
the flaring episode turns out to be 
$ N_{e-/e+} = \int dV \, (dn/d\gamma) \, d\gamma =2\times  10^{42}$, 
where $V$ is an assumed spherical volume of radius $L$.

\begin{figure}
\begin{center}
\includegraphics[width=13cm]{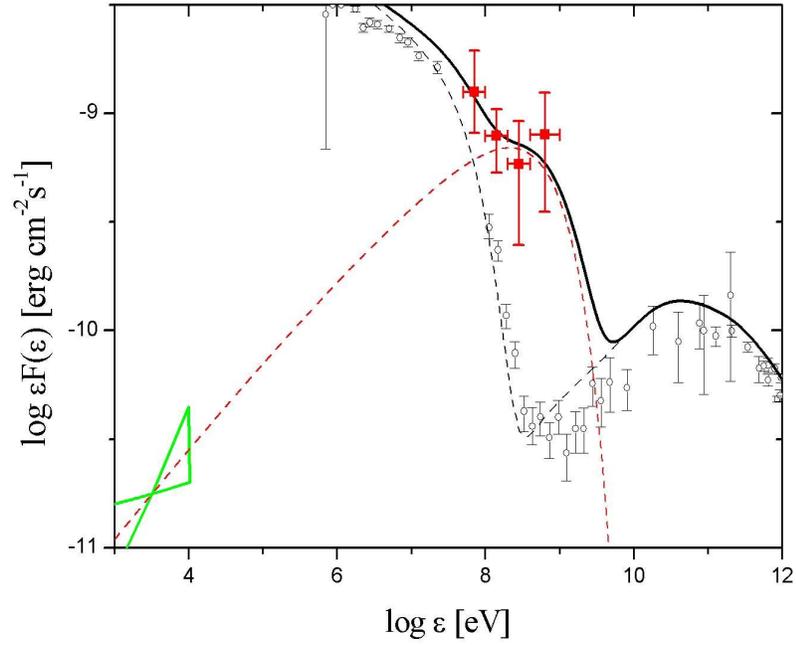}
\caption{AGILE (filled squares) 2-day averaged data of the Sept.
2010 gamma-ray flare of the Crab Nebula. Pulsar data have been
subtracted. The solid line represents the 2-day averaged
synchrotron emission model of the Sept. 2010 flare summed with the
standard nebular emission as discussed in the text. The dashed
curve in red shows the flaring component averaged over 2 days.
Data points in open circles give the standard average Crab Nebula
spectrum that we model by the dashed black curve. {\mt The
spectral region marked in green shows the X-ray data of "source A"
of T11}.} \label{fig-1}
 \end{center}
 \end{figure}

\begin{figure}
\begin{center}
\includegraphics[width=13cm]{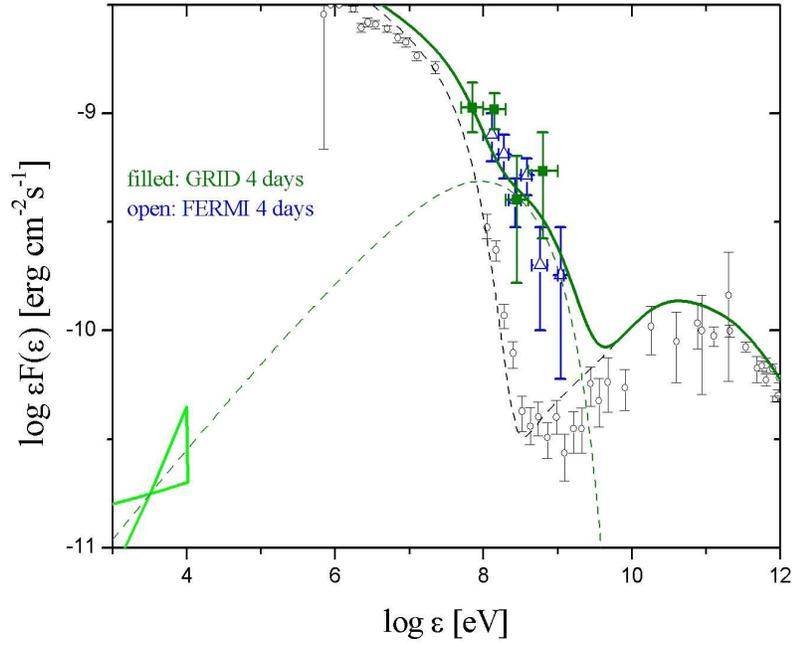}
\caption{AGILE (filled squares) and Fermi-LAT (open triangles)
4-day averaged spectral data of the Sept. 2010 gamma-ray flare of
the Crab Nebula. Pulsar data have been subtracted. The solid line
represents the 4-day averaged synchrotron emission model of the
Sept. 2010 flare summed with the standard nebular emission as
discussed in the text. The dashed curve in green shows the flaring
component averaged over 4 days. Data points in open circles give
the standard average Crab Nebula spectrum that we model by the
dashed black curve.} \label{fig-2}
 \end{center}
 \end{figure}

\begin{figure}
\begin{center}
\includegraphics[width=13cm]{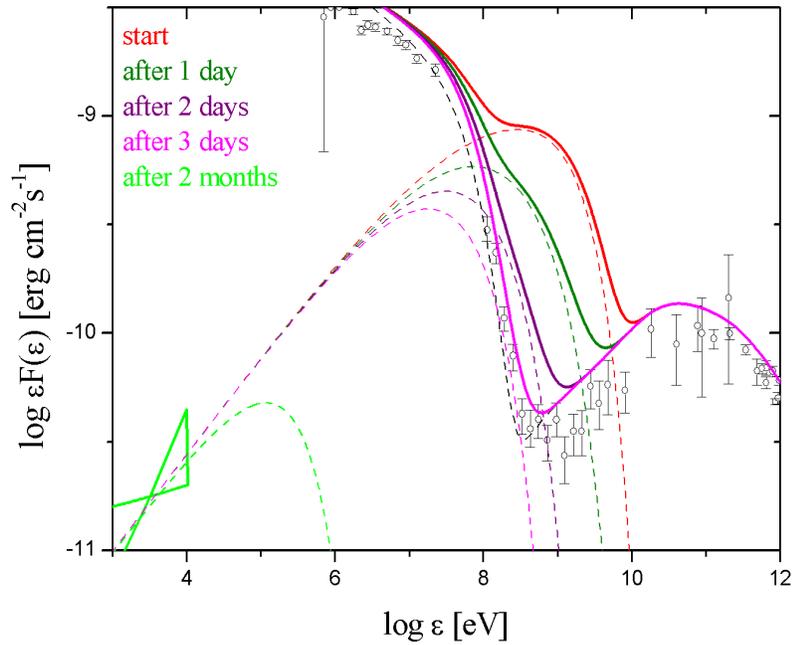}
\caption{Spectral evolution of the gamma-ray Crab Nebula September
2010 flare as obtained by our fast-rise-synchrotron-cooling model.
The upper curve shows the spectrum at the starting time, and the
lower curves show the spectra after 1, 2, 3, and 60 days
respectively. {\vv Note the persistence of the X-ray emission in
the "source A" of T11 localized by \textit{Chandra}.}}
 \label{fig-3}
 \end{center}
 \end{figure}

Based on standard synchrotron emissivity and particle cooling, we
calculated both the particle distribution and the photon spectrum
evolution keeping $B_{loc}=1$mG constant. We show in Fig.
\ref{fig-3} the calculated photon spectra at four different times
corresponding to days 1-2-3-4. Given our model parameters, fast
spectral evolution takes place, and the flaring phenomenon {\mt
fades} away within the fourth day. We also calculated time
spectral averages of the differential gamma-ray energy flux
$d\bar{F}/dE$ for different integration time durations $T$
according to the formula
\begin{equation}
\frac{d \bar{F}}{ dE } = T^{-1} \int_0^{T} \, \frac{d \bar{F}}{ dE
} \, dt \label{eq-3}
\end{equation}
with the particle energy loss rate $\dot{\gamma} = -
\gamma/\tau_s$ where $\tau_s$ is the synchrotron cooling time. We
use the time-integrated spectral function of Eq. \ref{eq-3}  to
model the 2-day (Fig. \ref{fig-1})  and 4-day (Fig. \ref{fig-2})
integrated spectral data of AGILE and Fermi-LAT. We find that the
synchrotron peak photon energy during day no. 1 is \be E _{peak }
=  \frac{3}{2} \, \hbar \, \frac{e \, B_{loc}}{m_e \, c} \,
\gamma_{max}^2 \simeq 800 \; \rm MeV   \label{eq-4} \en \noindent
which is in good agreement with the peak shown in the 2-day
averaged AGILE spectrum (Fig. \ref{fig-1}).
In the absence of very strong Doppler effects, our
measured spectrum and the calculated $E _{peak }$ violate the
expectations from Eq. \ref{eq-1}. Doppler effects with
${\cal D}\sim $ a few would not alter this conclusion. We find that the
emission from inverse Compton scattering of the flaring particle
population is negligible.

We also note that the spectral shape calculated in Fig.
\ref{fig-3} and measured in Fig. \ref{fig-1} contradicts a simple
translation by a Doppler factor of the average nebular data
showing the synchrotron burn-off. The additional population of
energized particles necessary to explain the flare can be
successfully modelled by Eq. \ref{eq-2} (that can also account for
the X-ray and optical "afterglow" in the anvil region measured by
\textit{Chandra} and HST, T11). 
Our model applies to emitting regions idealized as standing sites or
as regions within MHD outflows (such as the anvil features). 
Adiabatic expansion could play a role in contributing to
the gamma-ray flux decrease for an emitting site speed of order of the 
sound speed. We checked the relevance of the adiabatic expansion in our 
model and concluded that it could explain a good fraction of the observed  
flux decrease. In this case, our estimate of the the local magnetic field 
$B_{loc}$ would be an upper limit. Moreover, there would be no direct relation 
between the gamma-ray emission and the X-ray emission of the 
anvil enhancement because, at variance of the synchrotron cooling, 
adiabatic models cannot account for the persistent brightening in the 
X-rays (see Fig. \ref{fig-3}).

We note that also a purely Maxwellian
particle energy distribution (presented in T11, and resulting from
particle shocks with no non-thermal tails) can also in principle
explain the spectral evolution above 100 MeV: also in this
case there would be no direct spectral connection between the
gamma-ray emission and the X-ray/optical properties of localized
regions in the Nebula. 

\section{An even faster evolution model}

 In our study we considered also the possibility of a flux and
spectral evolution even faster than that shown in Figs. 1-3. The
analysis of the Fermi-LAT Sept. 2010 data by Balbo et al. 2010
suggests indeed that the spectral evolution may occur on an
overall timescale even faster than 1-2 days. A cooling
timescale of $\sim 1$day is explained in our synchrotron model for
$ B_{loc} \sim 2.5$ mG, and $\gamma_{max} = 5 \, 10^9$. In
this case, the overall flaring episode lasting $\sim 4$ days is
characterized by a sequence of short acceleration and cooling
episodes lasting 1-2 days. Our analysis remains valid also in
presence of a faster evolution. In this case, the magnetic field
is determined to have a value $B_{loc} \sim 10 \, \bar{B}$.

{\vv We note that slower events lasting 6-7 days, as the 2007
flare presented in (T11), can be interpreted by the same model
with larger region involved $L\approx 5\, 10^{16}$cm and similar
$B_{loc}$.}
Determining the gamma-ray temporal structure on timescales shorter
than 2 days is limited by photon statistics. The
time-resolved analysis of the AGILE gamma-ray data for the Sept.
2010 event will be presented elsewhere.

\section{Discussion and Conclusions}

The Sept. 2010 event of the Crab Nebula lasting $\sim 4\,
$days is currently the shortest detected gamma-ray flare. Our
analysis shows that the flux and spectral evolution of this event
are well described by a model characterized by very fast (shorter
than $\sim 1$day) particle acceleration and by synchrotron cooling
in a local magnetic field 5-10 times larger than the average
nebular value $\bar{B}$. {\mt Both} the AGILE and Fermi-LAT
gamma-ray spectral data {\mt are} consistent with each other
within a 4-day timescale. Our analysis of the AGILE data on a
2-day timescale clearly shows that the emission is peaked at the
photon energy of Eq. \ref{eq-4}, which is almost one order of
magnitude larger than the "synchrotron burn-off" constraint of Eq.
\ref{eq-1}. The flaring mechanism in the Crab Nebula is quite
remarkable: it  accelerates particles to the largest kinetic
energies (PeV) associable to a specific astrophysical source and
does it within the shortest time ever detected in a nebular
environment.

Our results challenge the physical assumptions underlying Eq.
\ref{eq-1} and in particular acceleration models based on "slow"
processes. As we showed above, explanations in terms of Doppler
boosting are problematic in light of the measured spectral
curvature of the AGILE data. Even though a theoretical study of
possible acceleration mechanisms consistent with the data
discussed here is beyond the scope of this paper, we can briefly
mention some of the difficulties. First-order Fermi acceleration
with particles gaining energy by diffusing stochastically back and
forth a shock front (e.g., Blandford \& Ostriker 1978, Bell 1978,
Drury 1983) appears to be too slow and {\mt is} drastically
challenged by our findings. In particular, it is difficult to see
how a diffusive shock acceleration mechanism can violate Eq.
\ref{eq-1}. A locally enhanced (over $B_{loc}$) electric field can
produce a sort of "runaway" of kinetic energy gains with an
acceleration rate larger than the synchrotron cooling rate.
However, despite some attempts and analogies with other
astrophysical contexts (e.g., pulsar magnetospheres),  it is
currently not clear how this mechanism can be implemented in the
Crab Nebula. MHD models of the pulsar wind (e.g., Komissarov \&
Lyubarsky 2004, Del Zanna et al. 2004, Camus et al. 2009,
Komissarov \& Lyutikov 2010), address the turbulence and the
limit-cycle behavior of the instabilities. These features may in
principle favor substantial local magnetic field enhancements.
However, the calculated timescales of these instabilities (e.g.,
Camus et al. 2009) are several orders of magnitudes longer than
what we detected in the Crab Nebula. Shock-drift acceleration
\cite{kirk} tends to occur on a timescale shorter than for
diffusive processes. However, it is not clear whether the required
efficiency can be reached in the flaring Crab Nebula site, and
whether Eq. \ref{eq-4} can be obtained. Shocks mediated by ions in
the pulsar wind that resonantly accelerate pairs by magnetosonic
waves \cite{gallant,spitkovsky,arons} are typically slow, and are
most likely not applicable in the X-ray and optically enhanced
pulsar polar jet regions of T11.

The challenge provided by the Crab Nebula gamma-ray flaring
requires a thorough investigation of the mechanisms leading to
efficient particle acceleration and to a natural justification of
Eq. \ref{eq-4}. The issue will be elucidated by future Chandra
X-ray and HST optical observations of the inner Crab Nebula that
will be carried out in search of the gamma-ray flaring site.

\noindent We thank an anonymous referee for his/her comments.

Research partially supported by the ASI grant no. I/040/10/0.



\end{document}